\documentclass[preprint]{elsarticle}

\usepackage{hyperref}

\journal{Computers \& Security}

\usepackage{soul,color}
\usepackage{hyperref}
\usepackage{booktabs}
\newcommand{\lmttfont}{\fontfamily{lmtt}\selectfont}

\usepackage{caption}
\usepackage{subcaption}

\usepackage{inconsolata}

\let\clsCenter\Center\let\clsendCenter\endCenter
\let\Center\undefined\let\endCenter\undefined
\usepackage{ragged2e}
\let\Center\clsCenter
\let\endCenter\clsendCenter

\usepackage{color}

\definecolor{pblue}{rgb}{0.13,0.13,1}
\definecolor{pgreen}{rgb}{0,0.5,0}
\definecolor{pred}{rgb}{0.9,0,0}
\definecolor{pgrey}{rgb}{0.46,0.45,0.48}

\usepackage{listings}
\usepackage{url}

\usepackage{threeparttable}

\usepackage{amsmath,amssymb,amsfonts}
\usepackage[super]{nth}
\usepackage{adjustbox}

\usepackage{etoolbox}

\makeatletter
\patchcmd{\ps@pprintTitle}
  {Preprint submitted to}
  {To appear on}
  {}{}
\makeatother

\patchcmd{\pprintMaketitle}
  {\ifvoid\keybox\else\unvbox\keybox\par\vskip10pt\fi}
  {\ifvoid\keybox\else\unvbox\keybox\par\vskip10pt\fi\ifvoid\extrainfobox\else\unvbox\extrainfobox\par\vskip12pt\fi}
  {}{}

\newenvironment{extrainfo}
  {\global\setbox\extrainfobox=\vbox\bgroup\parindent=0pt }
  {\egroup}
\newsavebox\extrainfobox

\begin{document}

\begin{frontmatter}

\title{Timing Covert Channel Analysis of the VxWorks MILS Embedded Hypervisor under the Common Criteria Security Certification}

\author{Domenico Cotroneo}
\ead{cotroneo@unina.it}
\author{Luigi De Simone}
\ead{luigi.desimone@unina.it}
\author{Roberto Natella}
\ead{roberto.natella@unina.it}

\address{DIETI - Università degli Studi di Napoli Federico II, Via Claudio 21, 80125 Napoli, Italy}

\begin{abstract}
Virtualization technology is nowadays adopted in security-critical embedded systems to achieve higher performance and more design flexibility. However, it also comes with new security threats, where attackers leverage \emph{timing covert channels} to exfiltrate sensitive information from a partition using a trojan. 

This paper presents a novel approach for the experimental assessment of timing covert channels in embedded hypervisors, with a case study on security assessment of a commercial hypervisor product (\emph{Wind River VxWorks MILS}), in cooperation with a licensed laboratory for the \emph{Common Criteria} security certification. Our experimental analysis shows that it is indeed possible to establish a timing covert channel, and that the approach is useful for system designers for assessing that their configuration is robust against this kind of information leakage.

\end{abstract}




\begin{extrainfo}
DOI: \url{https://doi.org/10.1016/j.cose.2021.102307}
\end{extrainfo}

\end{frontmatter}

\section{Introduction}
\label{sec:introduction}

\emph{Embedded hypervisors} bring virtualization technology from IT systems to embedded ones. Virtualization enables system designers to leverage modern high-end multicore boards, by partitioning resources among \emph{virtual boards} (i.e., an abstracted view of the underlying hardware board) \cite{sandstrom2013virtualization,aguiar2010embedded,mitake2011coexisting,heiser2008role,moratelli2016embedded}. This design approach comes with many benefits, including higher performance and efficiency, lower costs through off-the-shelf components and workload consolidation, and more flexibility and potential for innovation, by combining real-time OSes (e.g., for running control tasks) with feature-rich, general-purpose OSes (e.g., for running multimedia and HMI applications).

Since embedded systems are adopted in critical contexts, including automotive, aerospace, and industrial ones, they are increasingly exposed to security attacks, such as those reported by the press and researchers on SCADA systems \cite{langner2011stuxnet,ioactive2013compromising,tuptuk2018security}, satellite and aircrafts \cite{ioactive2014satcom,dessiatnikoff2012potential,costin2012ghost}, and top-of-the-line cars \cite{lemke2006embedded,checkoway2011comprehensive}. To mitigate risks, embedded hypervisors must prevent that compromised partitions (i.e., a general-purpose OS exploited by an attacker) could be leveraged to escalate privileges and to attack other partitions (e.g., the real-time, more critical ones). For this reason, security certification standards, such as the \emph{Separation Kernel Protection Profile} for the \emph{Common Criteria} certification \cite{ccportal,herrmann2002using,pfleeger2002security}, demand strong \emph{isolation properties} among partitions. Citing from the SKPP \cite{skpp}:

\begin{quote}
``The separation kernel allocates all exported resources under its control into partitions. The partitions are isolated except for explicitly allowed information flows. The actions of a subject in one partition are isolated from (viz., cannot be detected by or communicated to) subjects in another partition, unless that flow has been allowed. The partitions and flows are defined in configuration data."
\end{quote}

Due to the complexity of OSes and hypervisors, the certification process and, in particular, the assessment of isolation properties have proven to be a challenge, and have been limited so far to small systems at a high cost, e.g., using formal methods on a small subset of OS components \cite{heiser2008role,ge2019time}. Unfortunately, there is a lack of solutions for assessing these systems against \emph{timing covert channels}, where a compromised partition (e.g., running a trojan) modulates the use of a shared resource (such as, CPU and memory) such that another malicious partition (e.g., a spy) can infer bits of information from delays at accessing the shared resource \cite{lampson1973}. If an attacker succeeds in establishing a timing covert channel, it would be possible for him/her to have an unauthorized information flow between partitions, thus violating security policies and causing an information leakage.

In this work, we propose an approach for the assessment of embedded hypervisors against timing covert channels. We present the approach in the context of the security assessment of a commercial hypervisor product, the \emph{Wind River VxWorks MILS 3.0} \cite{windriver}, which we conducted in cooperation with a licensed laboratory for the \emph{Common Criteria} certification. Our approach includes attacks to the system call interface of the hypervisor, in order to force delays in the context switches between virtual boards, such that a malicious virtual boards can sense the presence of a delay (thus, one bit of information) through a statistical analysis of time measurements. Experiments clearly show that it is possible to establish a timing covert channel between virtual boards. We believe that the presented approach can be useful for system designers, to define an assessment method to verify the robustness of specific configuration to time covert channel threats.

The rest of the paper is structured as follows, We briefly survey related work on covert channels in section~\ref{sec:related}, and we provide background on MILS system in section~\ref{sec:background}. Case study along with its security features are discussed in sec.tion~\ref{sec:review_design}, In section~\ref{sec:review_evidences}, we review the \emph{certification evidences} provided by the commercial vendor in support of isolation properties. The proposed attack for the assessment of \emph{timing covert channels} is discussed in section~\ref{sec:covert}. The experimental setup is discused in section~\ref{sec:emulation}, while results are analyzed in section~\ref{sec:exp_results}. Section~\ref{sec:conclusion} concludes the paper. 

\section{Related Work}
\label{sec:related}

Leakage channels are a major threat for multi-level secure systems, where host separate \emph{information domains} on the same system, allowing access by users with different authorizations to sensitive resources. This is the case, for example, of OSes and cloud virtualization infrastructures. Leakage channels enable an attacker to obtain sensitive data from a domain without being authorized. In particular, a \emph{covert channel} is an information flow where an attacker abuses a mechanism not intended for communication, in order to violate the information flow policies of a system. A covert channel involves the collusion among two domains, one acting as a sender and the other as a receiver. A typical case of the sender is a \emph{trojan}, i.e., code that runs in a trusted domain but that operates maliciously \cite{lampson1973} (such as, an unverified library, a third-party app, or a web browser plugin), which sends information to a \emph{spy} program that runs within an unprivileged domain.

\begin{figure}[htbp]
\centering
\includegraphics[scale=0.4]{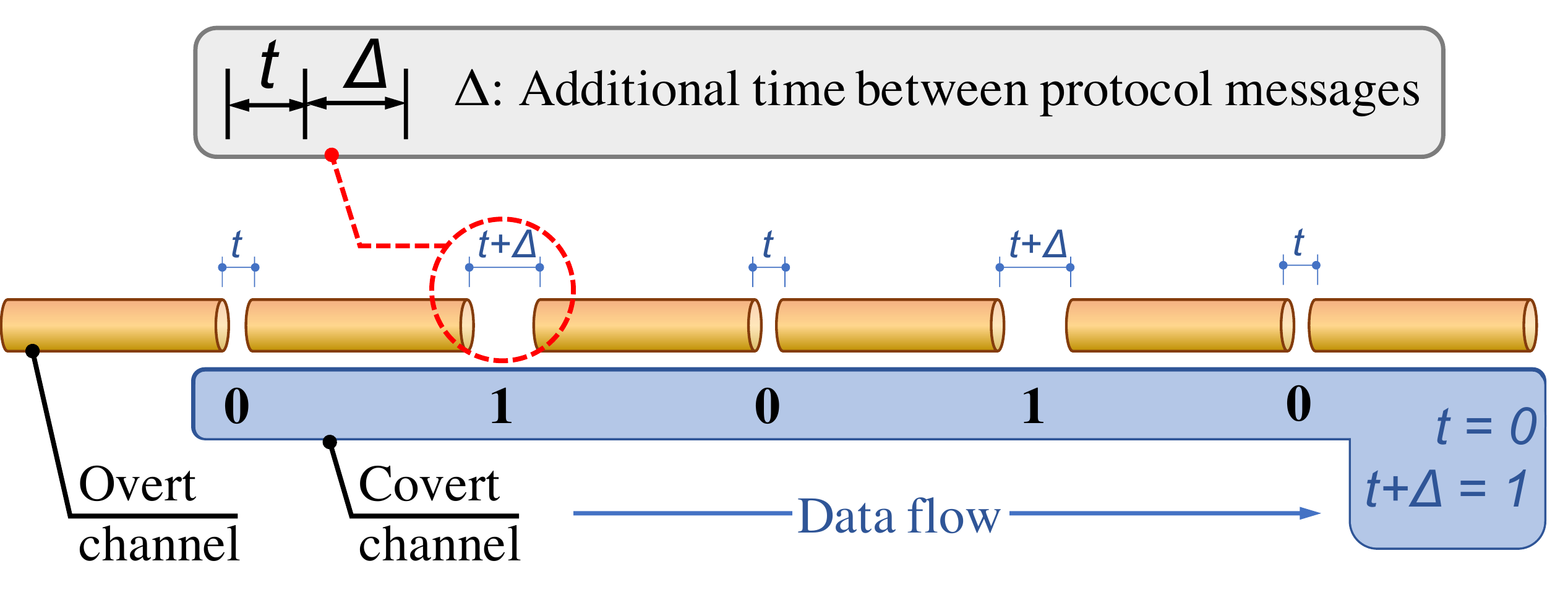}
\caption{Timing covert channel in a network environment.}
\label{fig:timing_covert_channel_net}
\end{figure}

Covert channels are usually classified into two types: \emph{storage channels}, where a process writes into a storage location (e.g., an unintended shared memory buffer) that is read by another process; and \emph{timing channels}, where a process sends information by modulating its usage of shared resources, in a way that influences the response time observed by the second process at accessing the resource. 
In a timing covert channel, the sender produces a stream of bits by accessing a shared resource, respectively with or without a delay, which is sensed by the receiver. 
For example (\figurename~\ref{fig:timing_covert_channel_net}), in a network environment, the attacker modulates packets over the network (the \emph{overt} channel) to introduce delays (the \emph{covert} channel). 

Research on timing covert channels has been  mainly focused on OSes, by addressing channels based on logical resources, such as the filesystem, by modulating operations for opening, reading, writing, and locking files \cite{kemmerer1983shared}, the TCP/IP network stack \cite{cabuk2004ip}, and internal kernel data structures such as the process table \cite{cioranesco2016process}. 
More recently, timing attacks have been targeting shared hardware, such as servers for public IaaS cloud computing, which run mutually-mistrusting virtual machines from different customers. These attacks force competition for bandwidth-limited hardware resources that are functionally transparent to software \cite{ge2018survey,ge2019time}.  Well-known attacks include \textsc{Prime+Probe} and \textsc{Evict+Time}, which establish a covert channel using CPU caches, by intentionally replacing certain memory blocks such that the spy can decipher the message bits based on the memory hit/miss latencies \cite{ge2018survey,ge2019time}. 
Research studies have been developing several proof-of-concept attacks, to demonstrate the feasibility of timing covert channels, by targeting: (1) \emph{microarchitectural state}, including data and instruction caches, TLBs, branch predictors, instruction- and data-prefetcher state machines, and DRAM row buffers \cite{aciiccmez2007predicting,hund2013practical,allan2016amplifying,gruss2015cache}; and (2) \emph{stateless interconnects}, including bus and on-chip networks, where concurrent accesses can cause a reduction of the available bandwidth \cite{hu1992reducing,woo2007analyzing,wu2014whispers}. 

The two main design approaches to prevent attacks at microarchitecture-level are: cleaning the state of resources during switches between domains, such as, by flushing CPU caches; and to partition resources, such as, by using \emph{page coloring} to prevent conflicts on the same cache lines by different users, and by disabling \emph{page deduplication} within the hypervisor. 
For example, Ge et al. \cite{ge2019time} recently proposed a new OS design that prevents timing covert channels, and that is amenable to formal verification, where the OS flushes on-core micro-architectural state on domain switches, with a deterministic latency, and where kernel code and data are partitioned, e.g., by creating separate copies of kernel code. 
Other approaches to prevent timing covert channels are to deliberately introduce noise in the system (e.g., by making system clocks fuzzy) to hinder attackers from inferring bits of information \cite{hu1992reducing,sethumadhavantimewarp}. 
All of these approaches may impose a high performance overhead: for example, flushing on domain switching is only feasible on L1 caches, as flushing other levels of caching leads to an excessive increase of cache misses and performance degradation. Moreover, these strategies may be inapplicable in some cases, because of the lack of hardware support, such as \emph{memory partition allocation} (MBA) technology in Intel CPUs that can only impose approximate limits on memory bandwidth utilization \cite{intel_mba}. 

Recent research has been developing solutions to achieve better trade-offs between security and performance. \emph{CC-Hunter} \cite{chen2014cc} is a solution for \emph{detecting}, rather than preventing, timing covert channels, by monitoring CPU events through a novel hardware framework (e.g., conflicts on cache lines, the integer unit, and the memory bus). Then, CC-Hunter detects bursts of conflicts over moving time windows, and performs time series analysis (e.g., by computing autocorrelation) to detect suspicious recurring patterns of the bursty periods. \emph{C2Detector} \cite{wu2014c2detector} detects timing covert channel attacks on the Xen hypervisor, by analyzing sequences of events using Markov and Bayesian models to detect anomalies. \emph{ReplayConfusion} \cite{yan2016replayconfusion} detects more sophisticated cache-based attacks, where the attacker varies the mapping of addresses, by recording and then replaying deterministically the application under monitoring, to detect differences in the cache miss rate due to changing mappings.

Compared to previous research, which proposed attacks and mitigations on individual resources at the OS or at the microarchitecture-level (e.g., cache, memory bus), this work proposes a new approach for creating timing covert channels at the system-level. Our approach stresses the hypervisor interface (hypercalls) on domain switches, to indirectly affect the OS (e.g., kernel locks) and microarchitecture state (e.g., kernel instructions and data inside caches), and to leverage delays in the time-between-context-switches, without relying on detailed knowledge of the attacker about the configuration of the system under attack. This kind of attack on kernel data and code has been hypothesized but never pursued so far by previous work \cite{ge2019time}. Moreover, while previous research mostly focuses on cloud computing environments, in this paper we consider timing covert channels in the context of a case study on an embedded hypervisor aimed at the Common Criteria security certification. Embedded systems represent an emerging application of virtualization technology, and pose specific requirements and restrictions that we address in this work.

\section{Background on MILS systems}
\label{sec:background}

The \emph{Multiple Independent Levels of Security/Safety} (MILS) is a high-assurance security architecture \cite{rushby1981design,gjertsen2008multiple}. 
The primary security function of a MILS is to \emph{partition} (or \emph{separate}) the subjects and resources of a system into security policy-equivalence classes, and to enforce the rules for authorized information flows between partitions.

The \emph{Wind River VxWorks MILS} is a commercial product that pursues these design principles. It supports information flow control, resource isolation, trusted initialization, trusted delivery, trusted recovery, and audit capabilities. The information flow policies are defined by the customer-defined \emph{configuration vector}. Wind River VxWorks MILS also includes the support tools and procedures used to accurately generate and securely distribute that configuration vector. Specific assurance requirements are allocated to those support tools and procedures. 

The Wind River VxWorks MILS provides a highly robust foundation for system services and applications in mission-critical embedded systems, and a high degree of assurance for the enforcement of related security policies. Such policies include those for the management of classified and other high-valued information, whose confidentiality, integrity and availability must be protected. For example, the VxWorks MILS separation mechanisms, when integrated within a high assurance security architecture, are appropriate to support critical security policies for the Department of Defense (DoD), Intelligence Community, the Department of Homeland Security, Federal Aviation Administration, and industrial sectors such as finance and manufacturing.

The Wind River VxWorks MILS complies with security requirements derived from the \emph{Separation Kernel Protection Profile} (SKPP) issued by the U.S. Government \cite{skpp,zhao2017survey}. The original goal of this product was to claim conformance with the SKPP. After the SKPP was sunset by the NIAP (\emph{National Information Assurance Partnership}) in 2011, Wind River decided to conform to a subset of SKPP assurance requirements that apply directly to the product that is provided by Wind River to its customers. This subset applies to generic Separation Kernel features, leaving out the features that are closely related to the hardware and firmware (whose responsibility is passed to the customer who builds the final system). 
The Separation Kernel features have been identified in the \emph{security target} of the Common Criteria certification, and are discussed in the following of this paper.

\section{Architectural review}
\label{sec:review_design}

The architect of an embedded system can design the overall MILS system by selecting \emph{virtual boards} (VB), by defining interactions between them (VB schedules, ports, and channels), and by creating a configuration vector. The configuration vector includes (but is not limited to):

\begin{itemize}
\item Virtual boards and images;
\item Communication channels between virtual boards (direction, mode, etc.);
\item Schedules for the virtual boards;
\item Authorized system calls for every virtual board;
\item Authorizations for memory-mapped I/O by virtual boards;
\item Physical-to-virtual interrupt mapping.
\end{itemize}

\begin{figure}[!ht]
\centering

\begin{subfigure}{0.5\textwidth}
  \centering
  \includegraphics[width=0.9\linewidth]{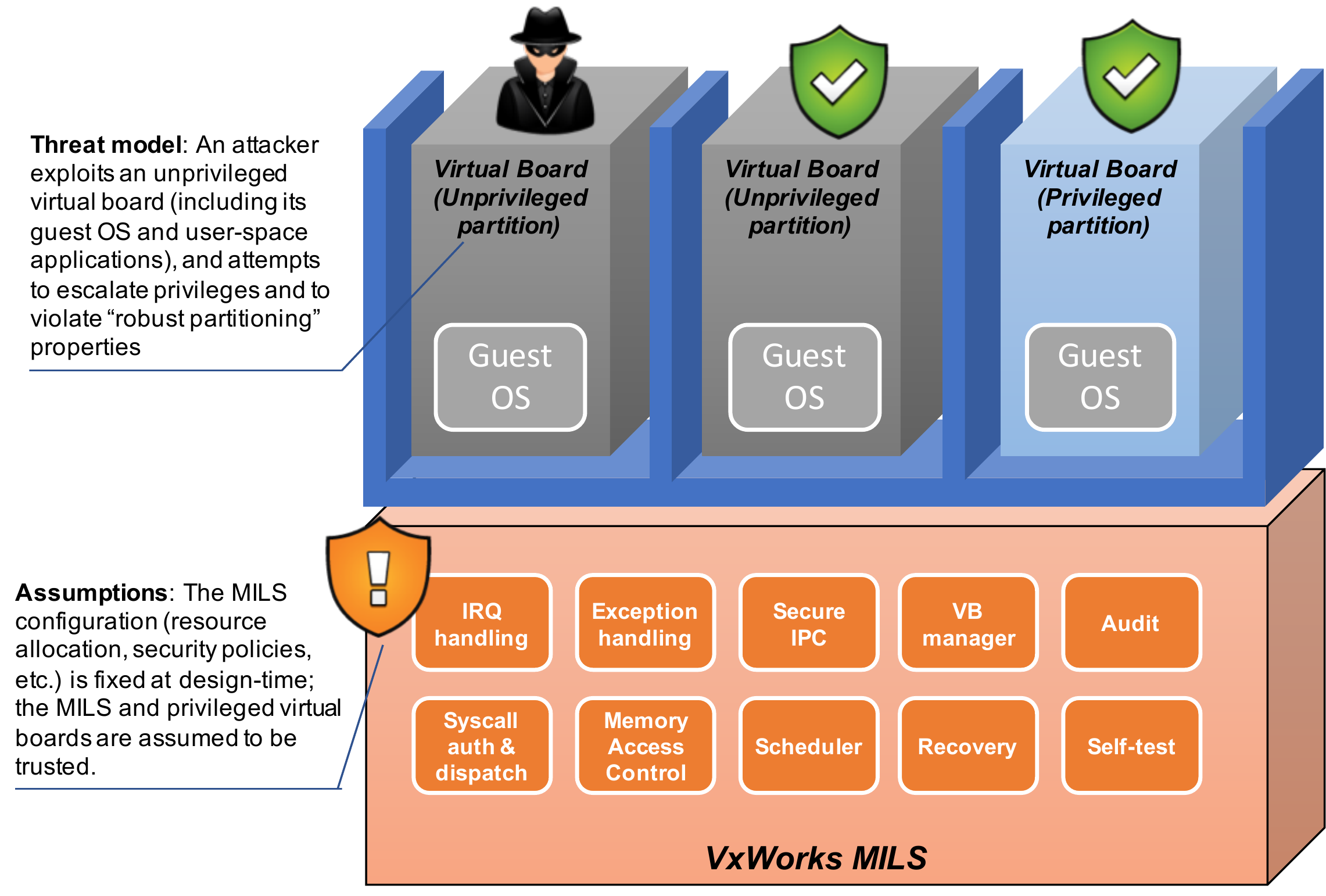}
  \caption{MILS threat model}
  \label{fig:mils_threat_model}
\end{subfigure}\begin{subfigure}{0.5\textwidth}
  \centering
  \includegraphics[width=0.9\linewidth]{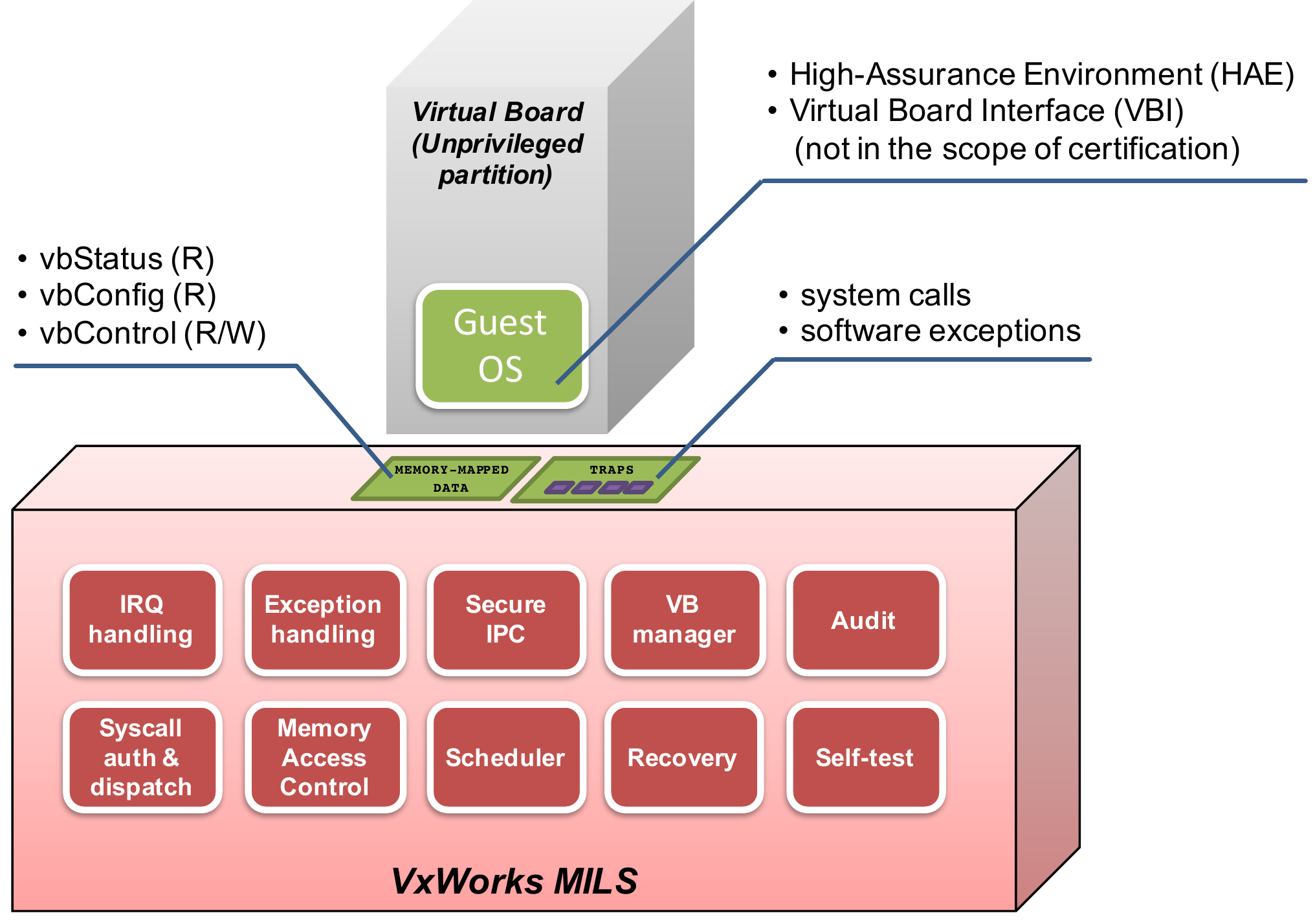}
  \caption{MILS attack surface}
  \label{fig:mils_attack_surface}
\end{subfigure}

\begin{subfigure}{0.5\textwidth}
  \centering
  \includegraphics[width=0.9\linewidth]{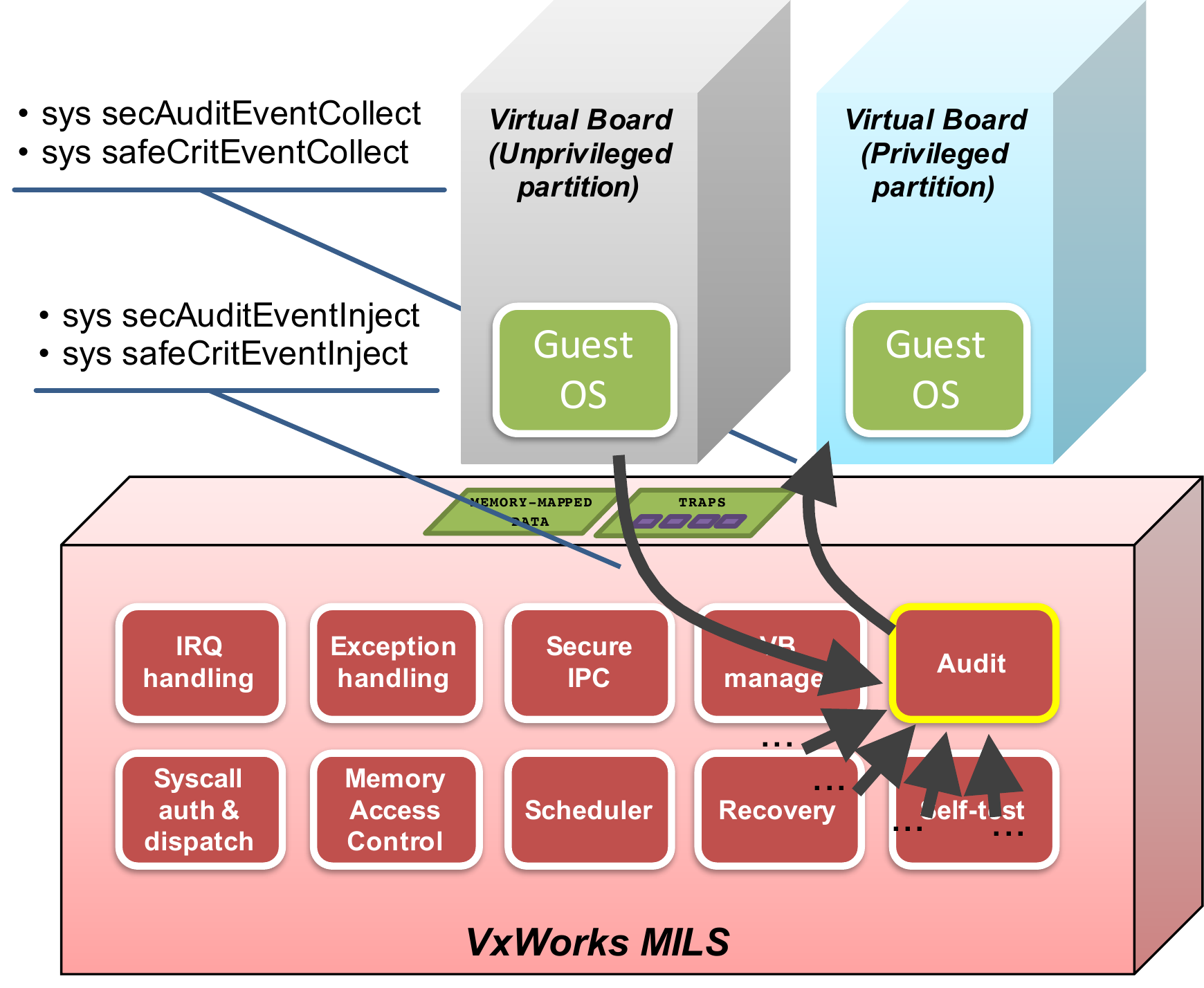}
  \caption{MILS security auditing}
  \label{fig:mils_audit}
\end{subfigure}\begin{subfigure}{0.5\textwidth}
  \centering
  \includegraphics[width=0.9\linewidth]{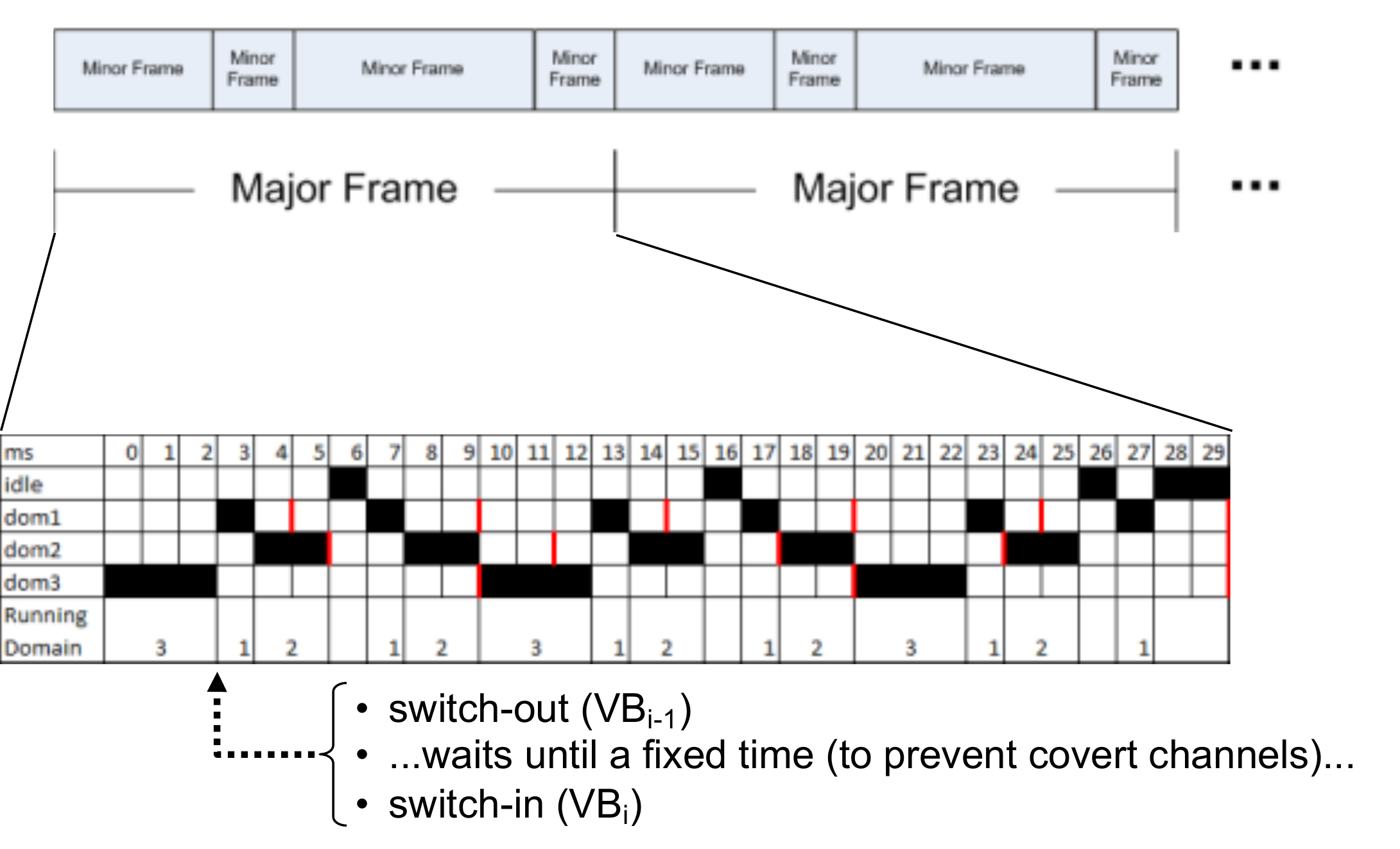}
  \caption{MILS scheduling (from \cite{xenarinc})}
  \label{fig:mils_scheduling}
\end{subfigure}

  \caption{Overview of MILS architecture and security aspects.}
  \label{fig:mils_design_review}

\end{figure}

The MILS establishes two primary security domains – MILS kernel (supervisor) space and virtual board (user) space. 
The MILS enforces a \emph{Least Privilege Abstraction Partitioned Information Flow Policy} (PIFP) to ensure security domains access only the resources that are required for their assigned functionality. The allowed information flows between specific VBs and resources are specified by the configuration vector and these flows are static. This architecture prevents that an attacker can exploit an unprivileged virtual board (including its guest OS and user-space applications) to escalate privileges and to violate \emph{robust partitioning} properties (Figure~\ref{fig:mils_threat_model}).

The MILS assumes \emph{protection from interference and tampering}, by protecting itself using a variety of methods:

\begin{itemize}

\item The MILS is established in a secure state at initialization time. The secure state is verified prior to authorizing any information flows. 
The MILS Payload BootLoader is the ``root of trust'' of the entire MILS system. This means that the MILS Payload BootLoader is implicitly trusted to be correct and cannot be tampered with. If the validation is successful, the MK initialization entry point is called. 

\item After initialization, the scheduler start function removes the MK initialization code and the part of the configuration vector that is only used by the MK for initialization. Then this function schedules the first VB for execution.
Most of the \emph{MILS Kernel} (MK) subsystems have no initialization function because each subsystem's data structures are created and initialized on the host. The only two areas that require runtime initialization are devices and MMU translation tables. 

\item At runtime, the MILS reference monitor and the Self-Test subsystem verify the MILS remains in a secure state.  If a failure causing the state to become insecure is detected, the MK Recovery is invoked to take action as specified by the configuration vector, which can result in rebooting or halting the system.

\end{itemize}

Moreover, the MILS assumes the \emph{non-bypassability of the MILS}. It ensures that \emph{Security Policy} enforcement functions are invoked and succeed before each function within the MILS scope-of-control proceeds. 

\begin{itemize}

\item Authorized virtual boards can only access the MILS through software exceptions -- directly via system calls or indirectly via non-system call exceptions.

\item The System Call Authorization and the Dispatch subsystem ensure only authorized MK system calls are allowed to execute, which is governed by the configuration vector.

\item If an illegal MK system call is invoked, a security audit event is injected and an error is returned to the caller.

\item Upon detecting a non-system call exception, the Exception subsystem injects a security audit event.  The action taken is specified by the configuration vector, which includes suspending the VB or forwarding the exception to the VB to be handled

\end{itemize}

The MILS Kernel exposes a minimalistic input surface to virtual boards, both in terms of quantity and complexity of its system calls. The input surface includes only 24 system calls across 10 subsystems, with three shared memory areas ({\lmttfont vbStatus}, {\lmttfont vbConfig}, {\lmttfont vbControl}) for exchanging information between a virtual board and the MILS Kernel (Figure~\ref{fig:mils_attack_surface}). These system calls include interrupt handling (masking, unmasking, acknowledging, setting the interrupt vector), memory management (updating the page table, flushing the TLB), and scheduling (switching schedule among a set of pre-defined configurations, donating time between virtual boards). Moreover, the MILS provides a subsystem and system calls recording safety- and security-critical audit events in the system (e.g., attempted violations of security policies), and for collecting these events in a privileged VB (Figure~\ref{fig:mils_audit}). CPU scheduling (Figure~\ref{fig:mils_scheduling}) is based on the ARINC 653 specification \cite{xenarinc}, where a fixed set of virtual boards is repeatedly scheduled with a fixed periodicity in a predictable way.

The input domain of system call parameters is fixed at design-time according to the static configuration, such as, the ID of resources (virtual boards, IPC channels, etc.) in input to the system calls. This input surface has not been designed for programming convenience (i.e., it does not provide a rich set of calls and parameters), but for minimizing the inputs and for reducing the potential inputs that could be attacked. To support application developers, the VxWorks MILS comes with a guest OS and a programming library, the \emph{High Assurance Environment} (HAE), which is built on top of system calls from the hypervisor and the guest OS to provide richer programming abstractions. Overall, the MILS architecture provides a solid foundation for the design of security-critical systems.
 
\section{Certification evidences}
\label{sec:review_evidences}

The VxWorks MILS comes with a package of \emph{certification evidences} for demonstrating robust partitioning properties (i.e., prevention of information leaks between unauthorized virtual boards). These evidences show that the MILS is secure in three ways: (i) \emph{by design}, where software vulnerabilities are prevented by adopting architectural solutions; (ii) \emph{by testing}, where the MILS is exercised with test cases to show that it can indeed tolerate attack conditions; (iii) \emph{by code inspection and review}, where the source code and architecture are analyzed by human reviewers to check that security properties are met. These certification evidences were developed by Wind River in cooperation with an external company (\emph{Verocel Inc.}), in order to guarantee the independence of V\&V activities from the original developers.

By design, the MILS architecture excludes that the system call interface can be circumvented. No MILS data is writable from userspace, and hardware MMU mechanisms guarantee the separation of memory address spaces. The MILS is single-threaded, and any attempt to flood the call stack is ineffective since the calling context is pended until the API call completes. Interrupt behavior is caged in such a way that servicing interrupts other than the timer interrupt is suspended for the duration of any system call. This mitigates any attempts to use interrupts to disrupt kernel processing.

By testing, the certification evidences show that system calls were correctly implemented, tracing them back to high-level requirements (in a semi-formal specification). Test cases include both \emph{functional tests} and \emph{robustness tests} (meant as a form of penetration testing), which are based respectively on \emph{valid} and \emph{invalid} input parameters. System calls were tested by defining \emph{equivalence classes} for every input parameter (e.g., attempt to pass malformed and invalid pointers, as well as pointers to areas that violate the security policy). The tests take advantage of white-box knowledge of the order of checks in the code to reduce the actual number of tests required to show complete coverage of valid and invalid input combinations.
The tests focus on unit-level tests for individual system calls and subsystems, but they deliberately leave out system-integration tests to the MILS customer. The tests assume that the MILS configuration does not impact on the robust partitioning properties. However, it is possible that an unusual MILS configuration (e.g., with a high number of VBs, IPC ports, short duration of major/minor scheduling frames, high number and size of memory partitions, etc.) may expose unexpected effects on CPU and memory management.

By code inspection and review, the certification evidences provide arguments that the MILS can achieve robust partitioning, using an assurance case in Goal Structuring Notation (GSN). These arguments focus on \emph{storage} and \emph{timing} covert channels, which were not covered by testing. A storage covert channel occurs when confidential information can be transmitted through a storage location (such as, a memory area or reserved bits of a header in a message) between users (in our context, the virtual boards) that are not allowed to communicate by the security policy. A timing covert channel occurs when users can transmit confidential information by modulating the use of system resources (e.g., CPU and memory), to change the response time observed by another user.

As evidence against storage covert channels, the vendor reports on an extensive manual analysis of MILS kernel code. For every system call, the reviewers identified every potential control flow path, using an ad-hoc static analysis tool, and enumerated all kernel variables read or written along these paths. The reviewers manually inspected every access to these variables, and provided arguments that the variable could not retain any information of the invoking virtual board that could be retrieved by another virtual board (e.g., by invoking another system call that reuses the same memory location). 
As evidence against timing covert channels, the vendor introduced a robust CPU scheduling mechanism to prevent that a virtual board could delay a context switch, so that another virtual board could receive unauthorized information by noticing the presence or absence of the delay. The MILS CPU scheduler enforces a fixed time-to-context-switch, by introducing an artificial delay at the end of the context switch, so that the time perceived by the virtual board is fixed (regardless of the time actually spent for the context switch). 

Even if the MILS vendor provides sound by-design and by-review arguments against covert channels, these arguments are not backed-up by tests, since testing for covert channels was (and still is) an open research challenge at the time the MILS was developed. The lack of automated verification represents a threat since manual reviews are prone to human error, and are very costly to replicate and difficult to use in a certification case. Additional tests are needed to identify covert channels: 
in this work, we propose a testing method to assess that the time-for-context-switch is indeed independent from the malicious activity of the virtual boards.
 
\section{Timing covert channel attack}
\label{sec:covert}

We developed a testing approach to assess the feasibility of timing covert channel attacks on the MILS kernel. The MILS is configured to run a mix of \emph{benign} and \emph{malicious} virtual boards, where the malicious one tries to establish a timing covert channel despite they are not authorized to communicate by any means. Our attack attempts from one malicious virtual board to delay the MILS kernel at domain switching, and from another malicious virtual board to sense the delay by sampling the system clock, and to perform statistical analysis on the samples to infer bits of information.

In the first malicious virtual board, we program a virtual interrupt at the end of its scheduling frame (\emph{Time Slice End}), $1ms$ in advance before the virtual board is switched out. Similarly, in the second malicious virtual board, we generate a virtual interrupt at the beginning of its scheduling frame (\emph{Time Slice Begin}). Both these virtual interrupts are generated by the MILS kernel on request from the virtual boards, and are documented features of the MILS that are normally available for architects of embedded systems.

\begin{figure}[!htb]
\centering

\begin{subfigure}{\textwidth}
  \centering
  \includegraphics[width=0.9\linewidth]{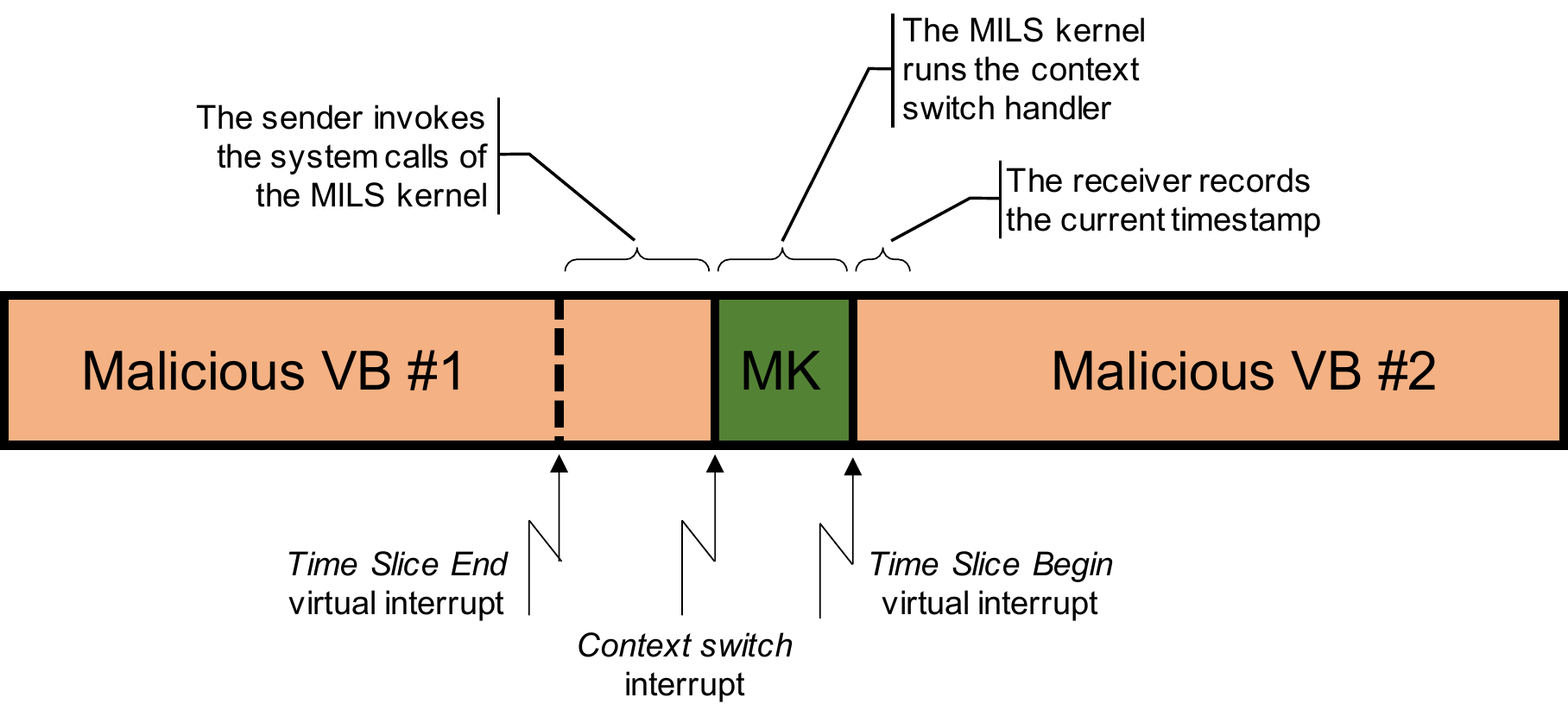}
  \caption{Context switch handling}
  \label{fig:timing_attack_interrupts}
\end{subfigure}

\vspace{5pt}

\begin{subfigure}{\textwidth}
  \centering
  \includegraphics[width=\linewidth]{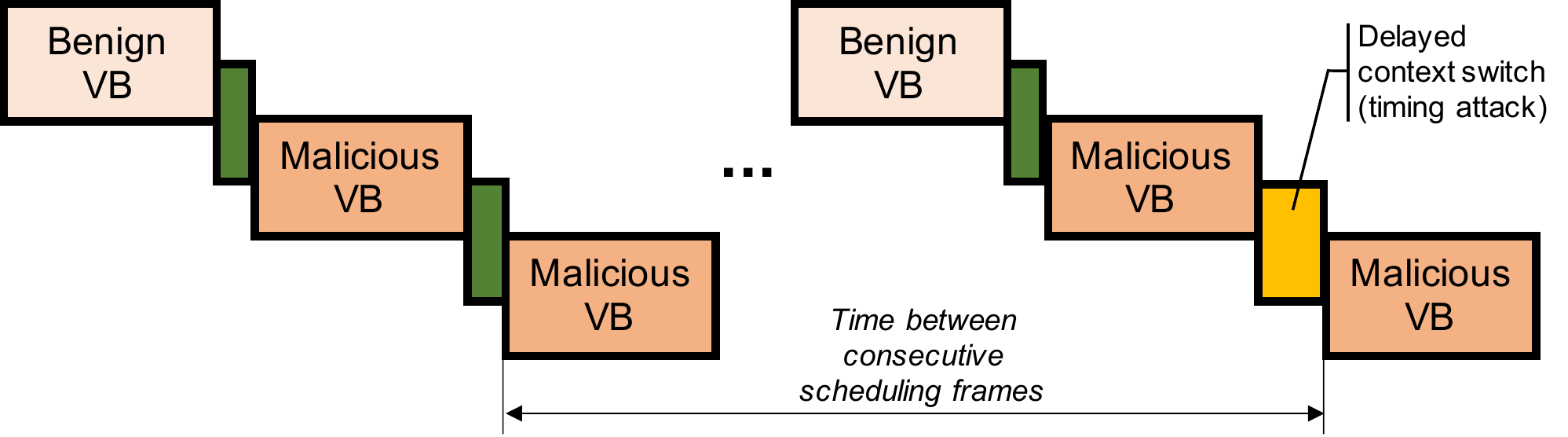}
  \caption{CPU scheduling under attack}
  \label{fig:timing_attack_schedule}
\end{subfigure}

\caption{Timing covert channel attack scenario.}
\label{fig:timing_attack}
\end{figure}

We use these programmed virtual interrupts to perform malicious activity before a switch-out, and after a switch-in (Figure~\ref{fig:timing_attack_interrupts}). Before the switch-out, our first virtual board performs system calls towards the MILS kernel (\emph{hypercalls}), so that the execution of a system call overlaps with the execution of the MILS scheduler and of the context switch handler. Our idea is that stressing hypercalls allow an attacker to trigger delays in the MILS kernel, e.g., due to flushing kernel code and data, and due to locking and interrupt masking, in a controlled way. The second virtual board records the current system time when the switch-in is completed. These timestamps are collected at each context switch of the malicious virtual board. For experimental purposes, we also log the timestamps through the serial port.

The attacker analyzes the timestamps to measure the time between consecutive scheduling frames of the malicious virtual boards. To prevent timing covert channels and assure robust timing partitioning, the MILS kernel must strictly follow a deterministic, periodic CPU schedule (see also Figure~\ref{fig:mils_scheduling}), regardless of malicious activities inside the virtual boards. We assess this property by evaluating whether the time-between-context-switches of the malicious virtual board is constant; otherwise, if the virtual board notices delays in the schedule, it could use this delay to infer confidential information from the other virtual board (Figure~\ref{fig:timing_attack_schedule}).

\begin{figure}
  \centering
  \includegraphics[width=0.9\linewidth]{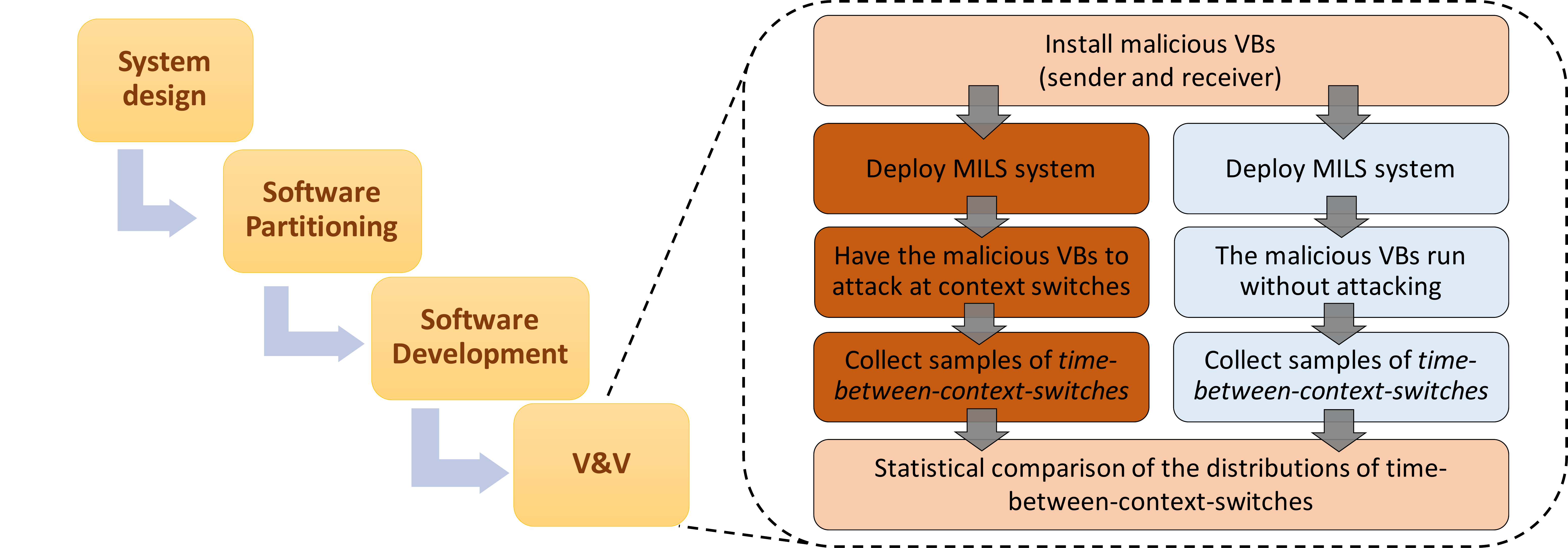}
  \caption{Workflow of the proposed approach}
  \label{fig:workflow}
\end{figure}

\figurename{}~\ref{fig:workflow} frames the attack in the context of a workflow for the system designer. The proposed approach can be applied during the V\&V stage of the product lifecycle, after the system designer established the configuration of the system, partitioned software functions across separated virtual boards, and developed the software. Then, as part of V\&V activities, the system designer wants to assess the robustness of the MILS system against timing covert channels. The system designer first installs the attack software, either in existing VBs or additional ones. Then, the MILS system is deployed and executed for two times, respectively \emph{with the attack} and \emph{without the attack}. During the executions, one of the VBs collects samples of the time-between-context-switches. Finally, the distributions of the samples from the two executions are compared by means of  hypothesis testing, in order to assess whether the attack has been able to cause a statistically-significant effect on the context switches.

The proposed approach makes little assumptions about the architecture of the system under assessment. In general, hypervisor-based embedded systems are designed to run mixed-criticality workloads, where low-criticality programs are deployed on distinct virtual machines from the high-criticality ones, and the hypervisor is configured to prevent any communication between them. In the proposed approach, we follow this general architecture, by having one virtual machine (representing the high-criticality part, where a trojan runs) that forces a delay in the hypervisor, so that another virtual machine (representing the low-criticality part) collects information from the observed delays. Other embedded hypervisors adopt a similar architecture, such as LynxSecure (which also complies to the MILS architecture), Xen (which can be adopted for embedded systems, such as in automotive projects), and seL4 (designed for high security guarantees through formal verification). For example, these hypervisors allow system designers to partition the resources of an embedded board among virtual machine, and make them to communicate only through secure channels. In a similar way to the VxWorks MILS, these hypervisors also perform deterministic scheduling in compliance with the ARINC 653 specification \cite{xenarinc}.

The proposed approach also makes little assumptions about which system calls have to be provided by the system under assessment. To apply the attack, the guest OS must be able to call system calls for reading the absolute time, also called ``wall clock time'' or ``real time'' \cite{vmware2008timekeeping}. In principle, it could be possible for hypervisors to prevent timing channels by denying attackers access to system calls for reading the real time, but in practice this is infeasible except in extremely constrained scenarios \cite{ge2019time,broomhead2010virtualize}. Thus, system calls for reading real time are generally available for the attack.

The approach also uses other system calls, but these are not mandatory for applying in the attack. We use two other system calls from the MILS: one to schedule virtual interrupts before the end of the time slice of the malicious virtual board, so that it can stress the MILS kernel while the context switch handler is invoked; and another to schedule interrupts at the beginning of the time slice, to measure the time-between-scheduling-frames. These system calls are not mandatory, since it is still possible (and confirmed by experiments in our testbed) to trigger the attack by invoking the MILS kernel throughout the duration of the time slice. This would make the activity of the trojan noisier, but would not prevent the success of the attack. Finally, our approach invokes system calls to stress the MILS kernel. We use memory-management system calls (such as \emph{vmmuConfig}) to increase the variability of MILS execution time, as memory-management configuration impacts on caching of kernel code and data, and uses locking and interrupt handling. Similar system calls are also available in other hypervisor products to allow the guest OS to manage virtual memory. Moreover, our approach is not restricted to use memory-management system calls, as any other system call which causes high overhead is also suitable for the attack.

The auditing mechanisms of the MILS are not meant to catch the timing attack, as the MILS only checks that the virtual board has permissions to call the system calls (which is indeed the case for our restricted subset of system calls). Detecting timing attacks at runtime is still an open research challenge. Section~\ref{sec:related} on discusses previous work on attack detection, but these solutions have not yet found their way into actual hypervisor products, as in practice they are quite difficult to deploy. Our proposed approach is also useful for assessing the capabilities of attack detection systems that may be proposed by future research.
 
\section{Testing environment}
\label{sec:emulation}

\begin{figure}[!htb]
\centering
\includegraphics[width=0.8\textwidth]{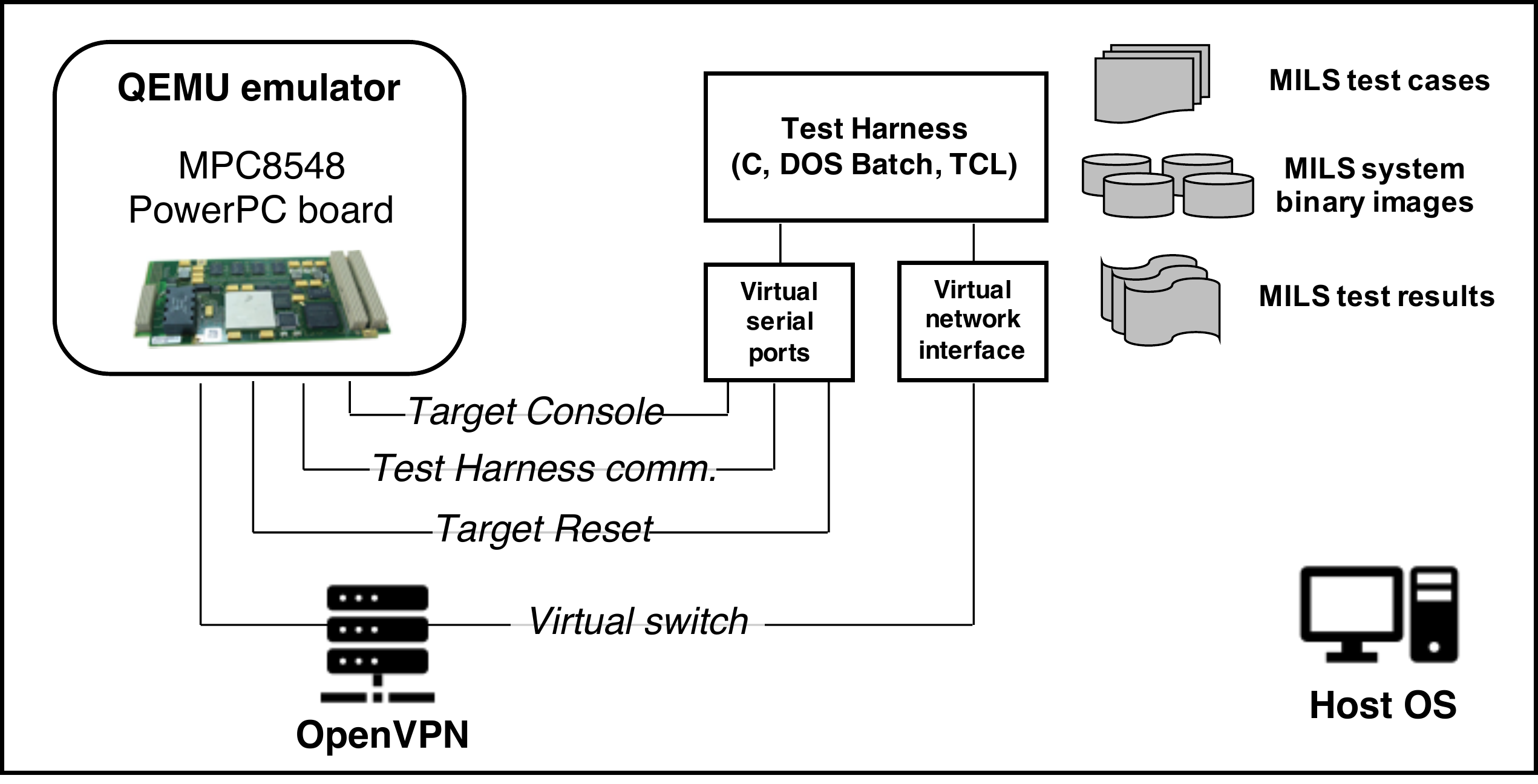}
\caption{Emulated MILS testing environment.}
\label{fig:mils_test_setup}
\end{figure}

To support verification activities by the certification laboratory team, and to experiment with the proposed approach, we prepared a testing environment based on a board emulator (Figure~\ref{fig:mils_test_setup}). Our environment reflects the testing environment used for certification by the MILS vendor: an MPC8548 PowerPC board \cite{powerquicc_cpu} running the MILS system, and connected to a host system through serial ports and a LAN. 
The emulated testing environment allows us and the certification laboratory team to perform more extensive and long-running tests, without wearing-out the physical hardware (e.g., prevent the wear-out of flash memory that could be caused by loading new code on the board). Moreover, the emulated testing environment allows us to experiment with covert channel attacks under different variants of the hardware configuration (e.g., under different CPU clock speeds).

The certification evidence package includes a set of test cases for the MILS, and a test harness for automating the built, execution, and reporting of the tests. Test cases are small C programs to be compiled and linked to the MILS system, which invoke the MILS system calls for functional and robustness testing. The test cases are designed to run on a set of four MILS system binary images, where each image is configured with a different set of fixed virtual boards with fixed IPC channels and shared memory areas. For every group of test cases, the test harness loads the appropriate MILS system binary image on the board; then, it sequentially uploads one test case on a shared memory area on the board, and runs the test.

Our emulated testing environment runs the MILS system on an emulated PowerPC board, based on the QEMU emulator (\url{https://www.qemu.org}), and with customizations by AdaCore (\url{https://github.com/AdaCore/qemu}) to emulate an MPC8548 board. 
We adapted the test harness to use emulated resources (e.g., a script for target reset was replaced by a restart of the emulator). 
Therefore, the emulated environment is able to successfully run the unmodified test harness and test cases from the certification evidence package.

We deploy a set of virtual boards, both benign and malicious, to attack the MILS kernel as described in Section~\ref{sec:covert}. The system configuration includes four virtual boards:

\begin{enumerate}
    \item A \emph{benign} virtual board ($B$), which runs a CPU-bound workload, on top of a Linux guest OS.
    
    \item A first \emph{malicious} virtual board ($M_1$), which simulates an attacker trying to create an unauthorized covert channel towards another malicious virtual board. It attacks the interface of the MILS kernel to delay the time-between-context-switches, in order to convey information by modulating the delays.
    
    \item A second \emph{malicious} virtual board ($M_2$), which receives data from the covert channel, by measuring the time-between-context-switches. 
    
\item A virtual board ($I/O$) that acquires the serial port from the MILS kernel, and collects data from the other virtual boards for debugging and experimental logging purposes, and transmits them through the serial port to an external PC. It receives data through authorized IPC channels, e.g., $M_2$ sends the time-between-context-switches samples for our analysis.
    
\end{enumerate}

We collect measurements of the time-between-context-switches from the virtual board $M_2$, in order to determine whether malicious virtual boards could affect context switches to establish a timing covert channel. Time measurements are collected by reading from the TBU/TBL performance monitoring registers, which are periodically increased by the CPU (\emph{ticks}) at the rate of a multiple of the CPU instruction clock \cite{powerpc_arch}. In the virtual board $M_1$, we schedule a \emph{Time Slice End} interrupt, and we run the attack in the interrupt handler, by repeatedly invoking the MILS system calls to force delays. We target memory-management system calls, such as {\lmttfont vmmuConfig} to reconfigure the page table and virtual memory and to force cache flushing, as these system calls have a greater overhead and performance impact compared to other, simpler system calls for IPC, event logging, or interrupt configuration.

\begin{figure}[htbp]
    \centering
\includegraphics[scale=0.3]{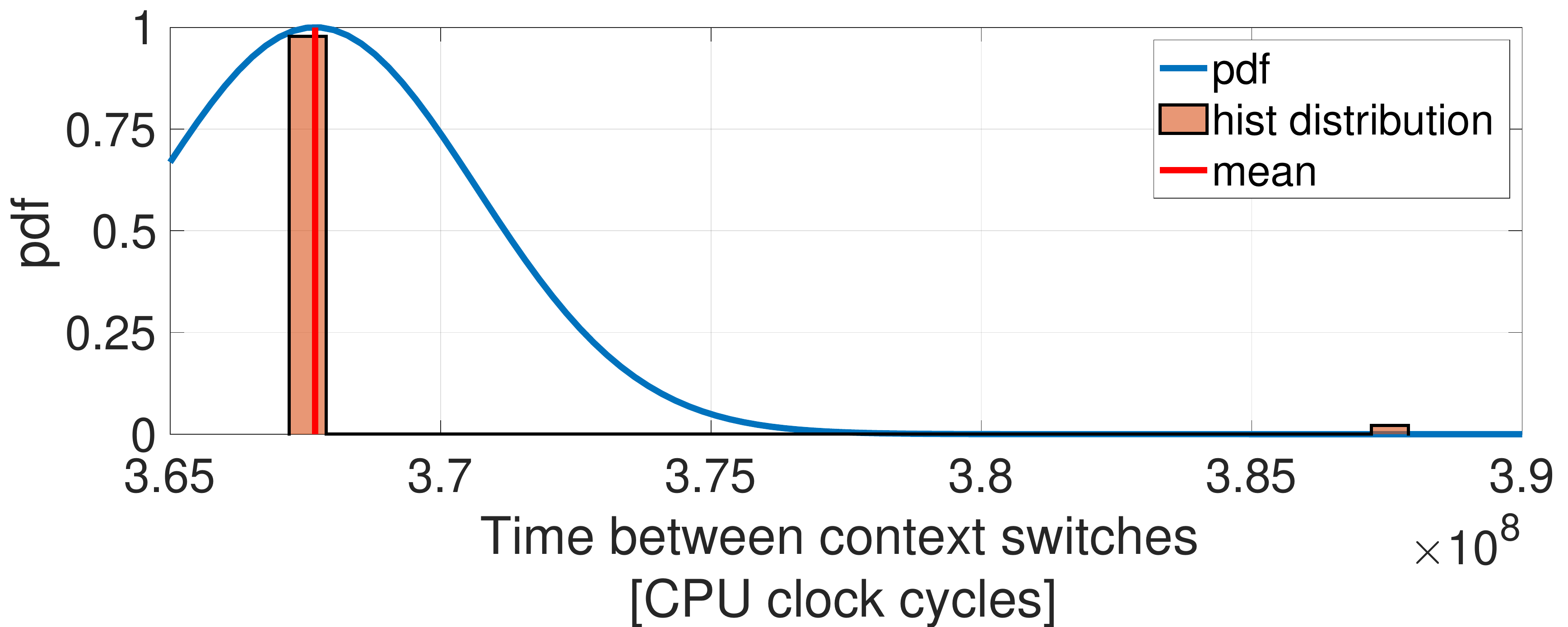}
        \caption{Time between context switches distribution against the MPC8548 physical board.}
        \label{fig:calibration_physical_board}
\end{figure}

Before analyzing timing covert channels, we perform a preliminary analysis and fine-tuning of the emulated testing environment, in order to make it representative of the real system running on a physical board, from the point of view of both functional and timing behavior. We built a MILS-based system image (i.e., an executable file, in ELF format, to be run on a board), by linking the MILS kernel to the images of the virtual boards. Then, we run it on an actual physical board, which is deployed in a licensed laboratory for the Common Criteria certification. 
The time slice duration of virtual boards has been set to $100 ms$, and the MILS timer OS tick frequency to $10 s^{-1}$.  Figure~\ref{fig:calibration_physical_board} shows the distribution of the measurements of the time-between-context-switches against the MPC8548 physical board. The distribution is tight around the average value, with only a few samples with higher values.

We also deployed and executed the same system image on the emulated testing environment, running on a machine with an Intel i7-4790 CPU (3.6GHz, 4 cores) and 16GB RAM, and with the Ubuntu Server 18.04 Linux OS. 
We configured the emulated testing environment to prevent that timing measurements could be affected by interferences due to CPU scheduling on the host machine (e.g., the emulator could be delayed due to CPU contention), by: \textit{i)} setting a higher priority for QEMU process on the host OS; \textit{ii)} by increasing QEMU process priority (i.e., using a lower \textit{niceness} value); \textit{iii)} by enforcing the execution of the QEMU process only on a specific, reserved physical CPU core, by \textit{pinning} the process; \textit{iv)} by removing unnecessary processes from the host machine. 
We compare the measurements of the time-between-context-switches from these two deployments to validate the emulated testing environment.

\begin{figure}[htbp]
        \centering
        \includegraphics[scale=0.3]{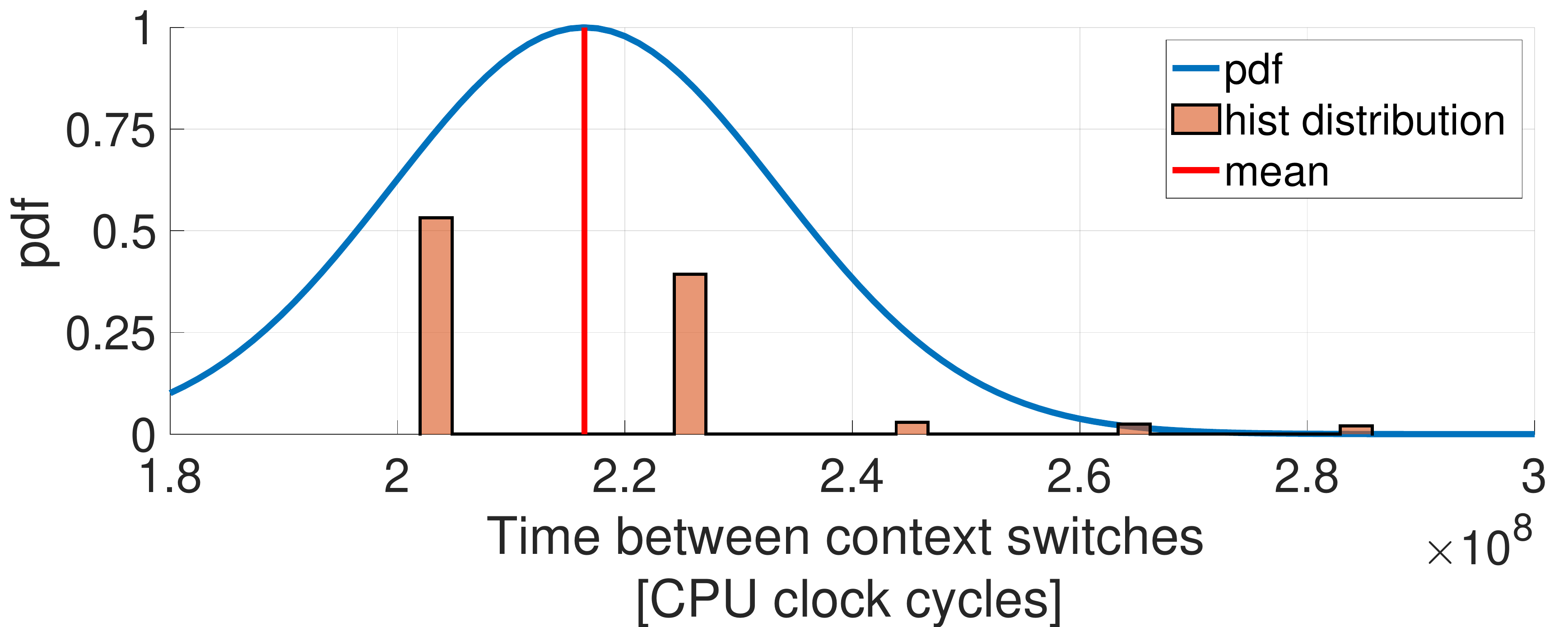}
        \caption{Emulated board, with host Linux OS tuning}
        \label{fig:calibration_qemu_tuning_icount0}
\end{figure}

We computed the distribution for the emulated environment, as showed in Figure~\ref{fig:calibration_qemu_tuning_icount0}. This distribution is bi-modal and clearly different from the one of the physical board. This motivated us to fine-tune the emulated environment to make it closer to the physical one.
By comparing the order of magnitude of the measurements ($3.6 \cdot 10^8$ CPU performance counter ticks in the physical board, vs. $2.2 \cdot 10^8$ CPU performance counter ticks in the emulated environment), we note that the virtual CPU runs at $3.6$Ghz, which is a higher frequency (i.e., amount of CPU ticks relative to wall clock time) compared to $1.5$Ghz of the physical board. This is motivated by the higher performance of the host machine compared to the embedded board. We executed the \emph{Dhrystone} performance benchmark \cite{weicker1984dhrystone} to confirm this behavior: the vendor of the board claims a score of $3,065$ \cite{powerquicc_cpu} (Dhrystone-converted MIPS value, v2.1, higher is better), while the benchmark on the emulator achieves a higher score of $12,110$.

To compensate for the performance difference and to reduce the gap between the emulated testing environment and the physical board, we further tuned the configuration of the emulator. We scaled the execution speed of the QEMU emulator, using the \emph{icount=N} parameter, where $N$ is a positive integer. The virtual CPU in QEMU executes one instruction every $2^N$ ns of virtual time; by increasing $N$, the execution speed becomes slower, in order to better align with the execution time of slow devices \cite{qemu_doc}. The previous experiment (\figurename{}~\ref{fig:calibration_qemu_tuning_icount0}) executed with $N=0$, which is the default value. Thus, we analyze the cases $\textit{icount}=1$ and $\textit{icount}=2$ to make the execution closer to the physical board.

\begin{figure}[h]
        \centering
        \includegraphics[scale=0.3]{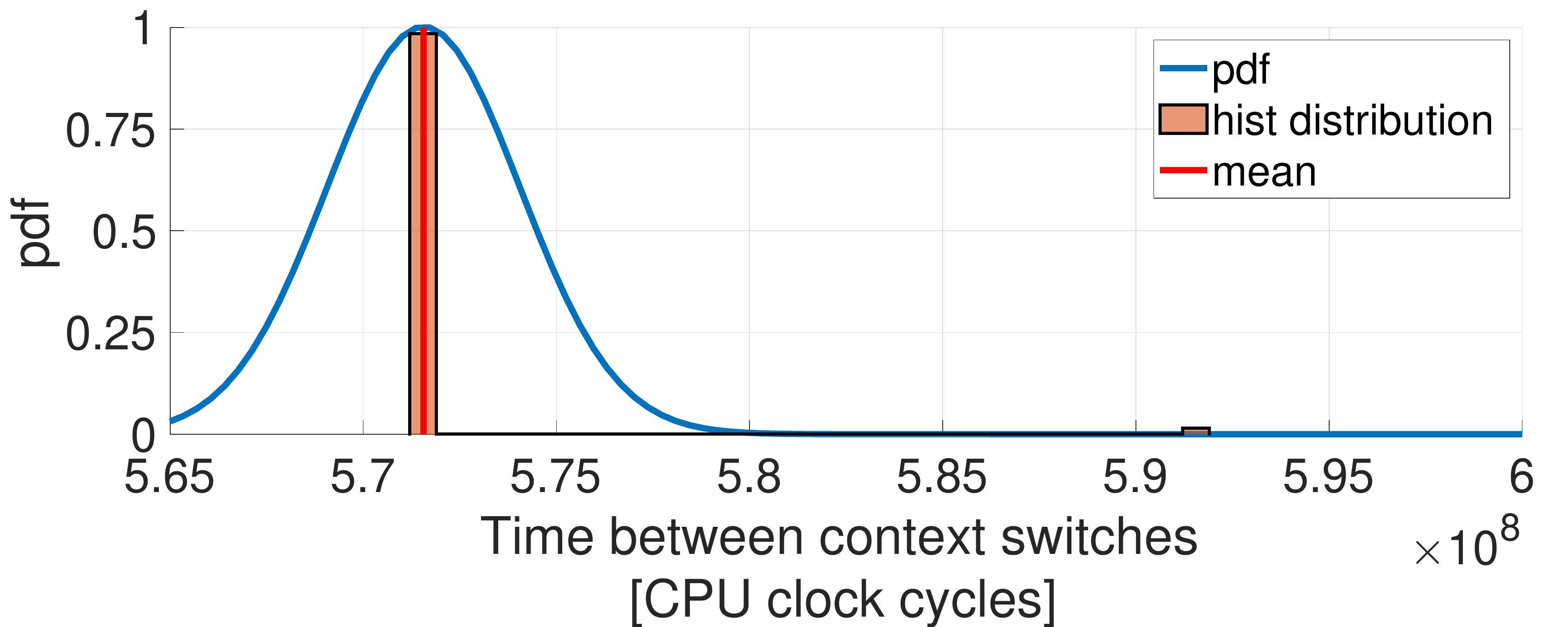}
        \caption{Emulated board, with virtual CPU speed scaling (\emph{icount=1})}
        \label{fig:calibration_qemu_tuning_icount1}
\end{figure}

\begin{figure}[h]
    \centering
        \includegraphics[scale=0.3]{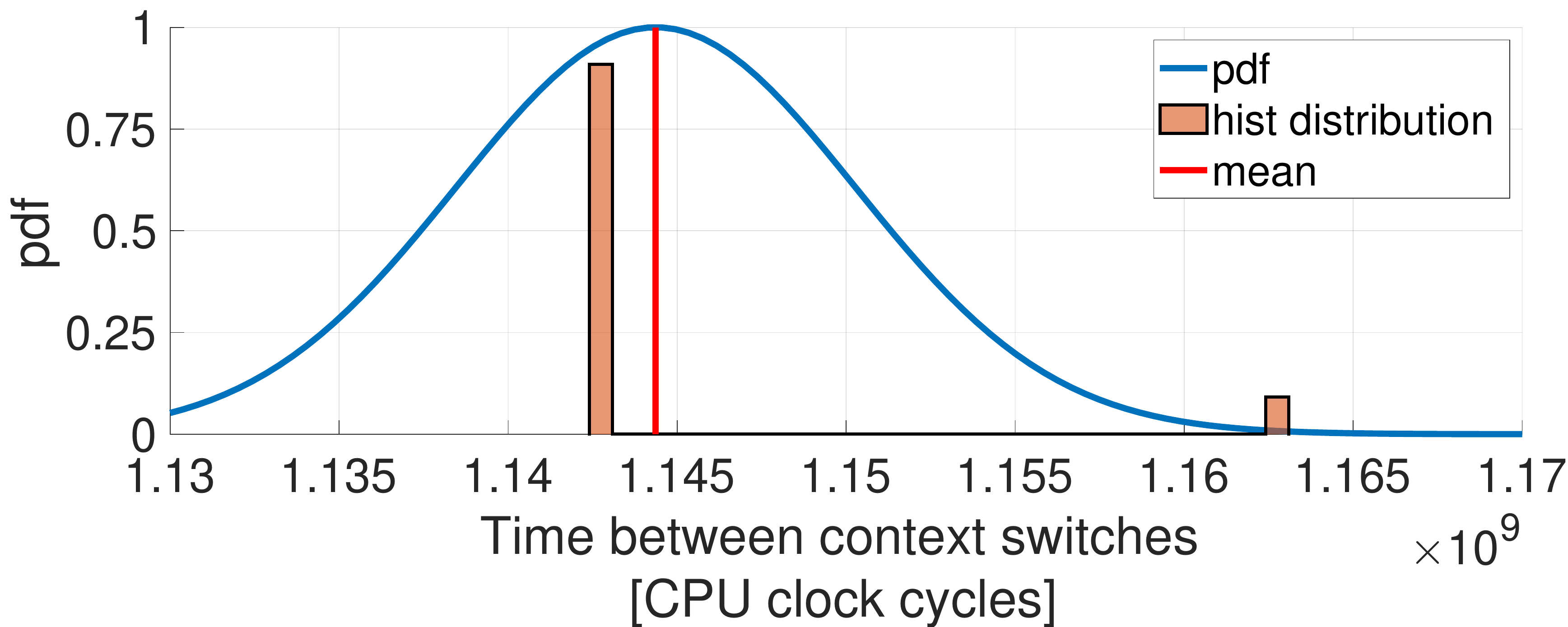}
        \caption{Emulated board, with virtual CPU speed scaling (\emph{icount=2})}
        \label{fig:calibration_qemu_tuning_icount2}
\end{figure}

\figurename{}s~\ref{fig:calibration_qemu_tuning_icount1} and \ref{fig:calibration_qemu_tuning_icount2} show timing measurements when the execution speed is slowed down. We note an increase of the order of magnitude, and that the shape of the distribution is similar to the physical board, where most of the samples are tight to the average. 

The execution with $\textit{icount}=1$ shows a behavior closer to the physical board than the run with $\textit{icount}=2$, which exceedingly slows down the execution. Therefore, in our experiments, we will assume that $\textit{icount}=1$ is the configuration most representative of the physical board. In our experimentation, we will tune the configuration of the environment, including the execution speed, to investigate the relationship between the configuration and timing covert channels.

\section{Experimental analysis}
\label{sec:exp_results}

The goal of our experimentation is to identify if, and under what conditions, attackers can establish a timing covert channel between unauthorized partitions, thus violating the timing isolation enforced by the MILS.

We designed an experimental plan to assess the influence of different factors (i.e., the configuration of the virtual boards and the MILS kernel) on the success of the attack. We introduce variations to the following parameters:

\begin{itemize}

    \item \textbf{Switch/Switch in/Switch out duration:} the duration of the context switch, where the MILS kernel intentionally adds a delay if the context switch lasts for less than this duration, to make it more deterministic. We set the switch duration of the malicious virtual boards to $10 ms$, $10$$\mu$$s$, or $1$$\mu$$s$.
    
    \item \textbf{Tick frequency:} interrupt frequency of the virtual timer (e.g., the periodic interrupt used by the Guest OS for CPU scheduling) in a virtual board, where all virtual boards have the same frequency. We set the frequency to $10 s^{-1}$ or to $1000 s^{-1}$.
    
    \item \textbf{Number of benign virtual boards:} we consider both the cases with only one benign virtual board $B$, and an extreme case with $50$ benign virtual boards.
    
    \item \textbf{CPU execution speed:} We scale the CPU execution speed by a factor $N$, using the \emph{icount} parameter as discussed in the previous section. We set this parameter to $N=0$ (a high-performance CPU), $N=1$, or $N=2$ (a low-performance CPU).
    
\end{itemize}

In total, we have $36$ combinations of these parameters. For each combination of these parameters, we perform both an experiment \emph{without} the timing attack (i.e., no malicious activity by the malicious virtual boards), and experiment \emph{with} the timing attack (i.e., a malicious virtual board stresses the MILS kernel system call interface to cause delays). Therefore, the total number of experiments is $72$. The experiments run the system for $15$ minutes, which allows us to collect enough measurements to analyze the time-between-context-switches with statistical methods. We repeat every experiment three times, to assess that the outcome is reproducible and statistically significant.

We perform a statistical hypothesis test (the \emph{t-test}) to assess whether there are statistically significant differences between the runs \emph{with} and \emph{without} the attack under the same configuration of parameters. We test the \emph{null hypothesis} that the distributions of the time-between-context-switches have the same mean value. The t-test points out whether the experiment \emph{with} the attack exhibits a delay with statistical confidence (i.e., the difference is unlikely to be due to random variations).

\begin{figure}
     \centering
     \begin{subfigure}[b]{\textwidth}
         \centering
         \includegraphics[scale=0.3]{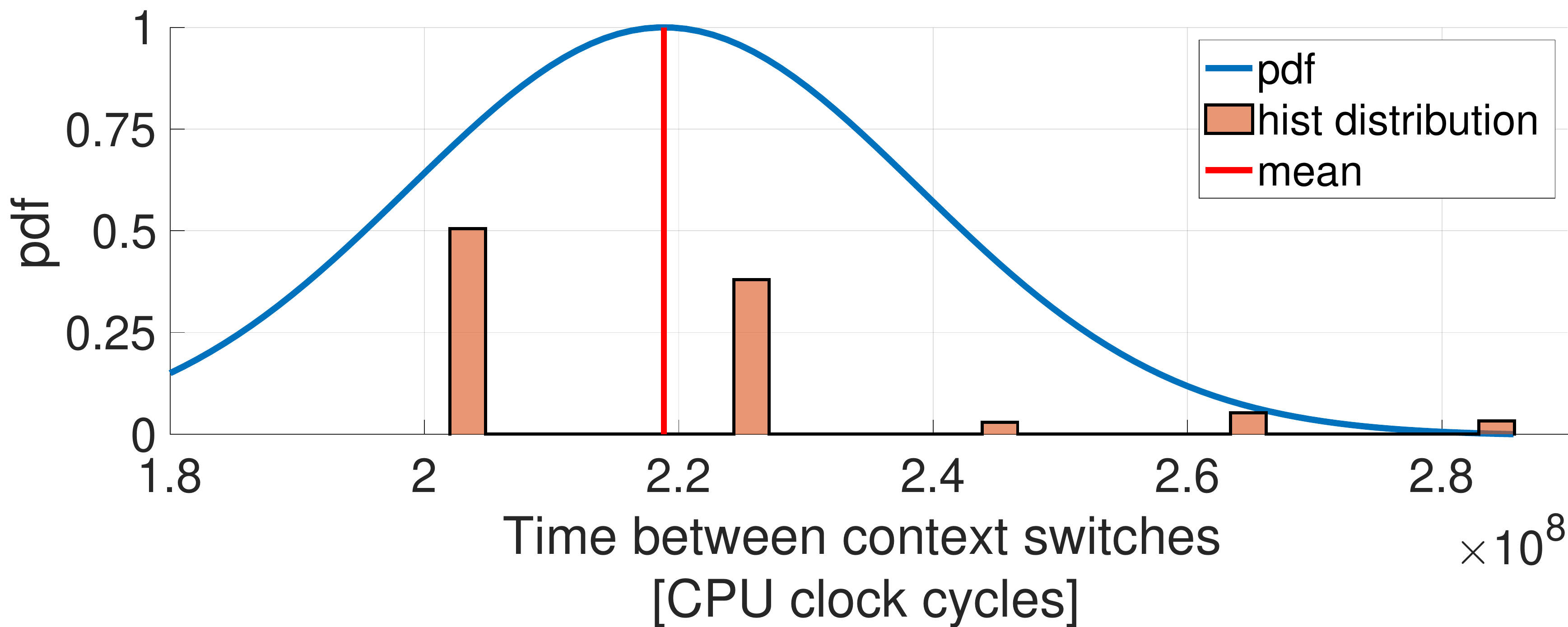}
         \caption{During no attack}
         \label{fig:example_attack_success_NO_ATTACK}
     \end{subfigure}
     \hfill
     \begin{subfigure}[b]{\textwidth}
         \centering
         \includegraphics[scale=0.3]{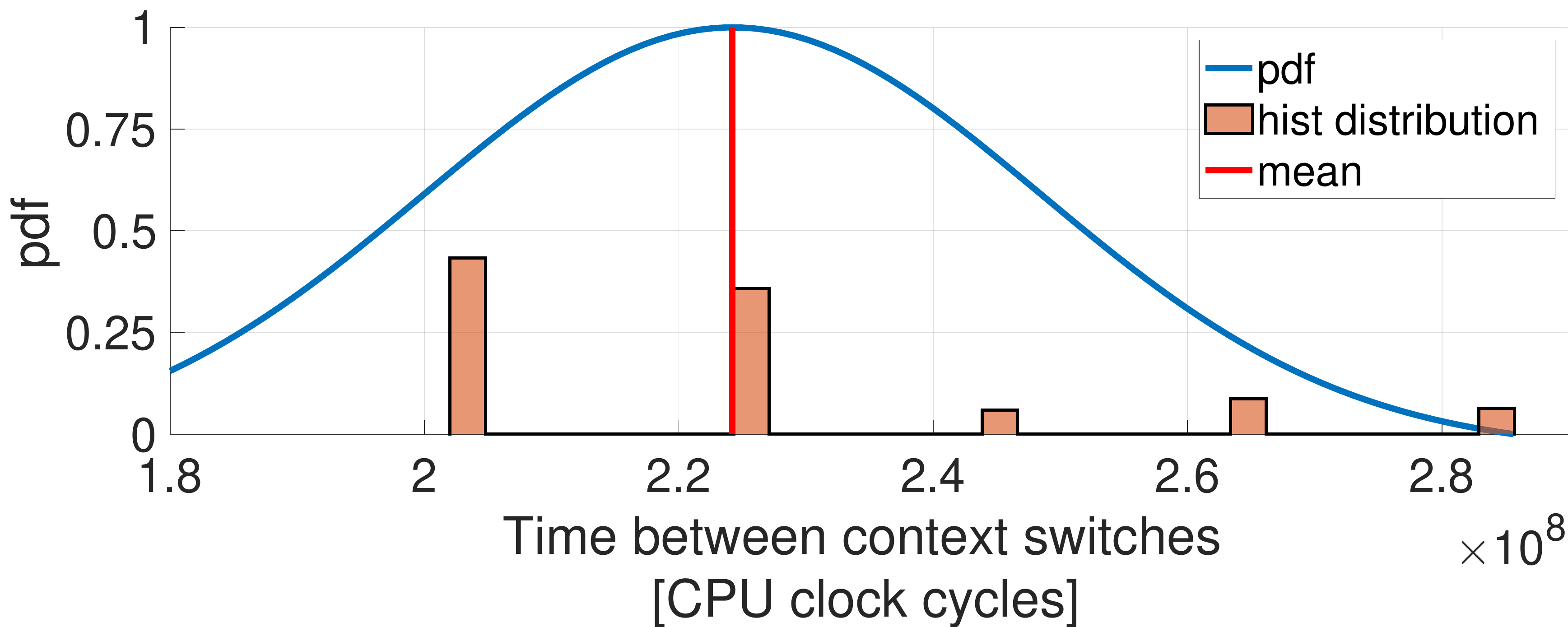}
         \caption{During attack}
         \label{fig:example_attack_success_ATTACK}
     \end{subfigure}
        \caption{MILS timing behavior under successful timing covert channel attack.}
        \label{fig:example_attack_success}
\end{figure}

From our experiments, we found that a \textbf{timing covert channel could be established when the execution runs at the highest speed ($N = 0$), and the context switch duration is set to $10$$\mu$$s$ or less}. In these configurations, we found a statistically significant difference, with the confidence of $95\%$, between the executions \emph{with} and \emph{without} attack. This outcome was reproduced across the repetitions. In the remaining configurations (e.g., $N = 1$ and $N = 2$), the executions \emph{with} and \emph{without} attack did not exhibit a statistically significant difference.

\figurename{}~\ref{fig:example_attack_success} shows an example of distribution of the time-between-context-switches in the case of a successful attack. In this case, there have been enough samples beyond the average value, which shifted the distribution towards higher values (i.e., higher bins on the right of the plot). According to the t-test, the difference is large enough that could not be simply attributed to random variations. This is confirmed by applying the t-test on two experiments both without the attack: in that case, this difference in the mean value could not be reproduced. Therefore, we can reject the null hypothesis and we can attribute the difference to the malicious activity of the virtual boards.

These artificial delays could be leveraged by a malicious virtual board to infer bits of confidential information over a long run (e.g., no delay $=0$, delay $=1$). Such an attack could be possible if the attacker has gained access to the virtual board, and if the virtual board is authorized to use the system calls for reproducing the attack. One possible reason for the vulnerability may be \emph{CPU cache flushes} made by the {\lmttfont vmmuConfig} system call, which could have slowed down the execution of MILS due to instruction cache misses. Another potential cause may have been critical sections in MILS code (e.g., code that disables interrupt handling to prevent preemption, or that uses semaphores for synchronization), which delays the context switch after the execution of the critical section.

In the lower-speed CPU configurations (i.e., $N = 1$ or $N = 2$), the attack does not succeed. We note that the higher-speed CPU ($N = 0$, in \figurename{}~\ref{fig:example_attack_success}) shifts the distribution of the time-between-context-switches to the left (i.e., lower absolute values), and causes a higher variance of the distribution. Under higher execution speeds, second-order factors, such as caching, have a higher influence on the time to execute MILS code, thus the higher variability, and the higher sensitiveness of the time-between-context-switches to the attack. Similarly, setting a higher context switch duration (i.e., artificial delays introduced by the MILS kernel to make context switch time deterministic) prevents the timing covert channel, as the additional delay compensates for the variability caused by the attack. We confirmed these behaviors by running the experiments on the physical board (which matches the experiments with $N = 1$): in the MPC8548 board, with that specific CPU, the MILS kernel can prevent the timing covert channel by introducing a long-enough context switch time delay.

These experimental results point out that timing covert channels may occur under specific hardware and software configurations. Therefore, it is important to execute tests as we performed by using our proposed approach, in order to assess whether a configuration is resistant against timing covert channels under the specific conditions, and to introduce proper mitigations. To prevent the attack, the MILS kernel should prevent untrusted virtual boards (e.g., running applications developed at a low integrity level, or exposed to the network) to use readings of the real system time from accurate sources (e.g., CPU performance counters); however, this solution could be excessively restrictive for implementing user applications. Therefore, the main mitigation against the attack is to configure CPU scheduling with a significantly long context switch duration delay, to force the MILS to wait for a longer ``slack'' time. The slack time increases the total execution time of the virtual boards (thus, slowing down the system), but also to absorb variations of the time-between-frames caused by the attack, thus reducing the variability of the distribution and making them indistinguishable from a normal condition, thus preventing transmission of information.

\section{Conclusion}
\label{sec:conclusion}

We presented an approach for the experimental assessment of isolation properties against timing covert channels, in the context of VxWorks MILS, a commercial hypervisor product for security-critical embedded systems. In this paper, we reviewed the architecture, the security features, and the evidence package for security certification. We found that the vendor performed extensive functional tests, and design reviews to support its certification arguments on robust partitioning; however, no tests were performed to assess that isolation properties actually hold against timing covert channels. Finally, we presented an experimental evaluation to show that timing covert channels are indeed feasible, and that system integrators need to take care of this risk with more restrictive security policies and CPU schedules.

The proposed approach can serve designers at assessing the robustness of their systems against timing covert channels. Of course, selecting a hypervisor with a security certification package, such as the VxWorks MILS, is a good choice, as it has been designed to reduce the attack surface and to thwart many attacks (mostly with respect to spatial isolation), and has been rigorously tested to achieve high confidence. However, the state-of-the-practice still falls short with respect to temporal isolation. Since timing covert channels are a concern for many scenarios, it is important to complement the certification package with additional tests, as in our proposed approach. Moreover, it makes sense for industrial actors to benchmark the timing isolation properties of competing hypervisor solutions (e.g., commercial against open-source ones). This is an important direction for future work that we are currently exploring.

\section*{References}

\bibliographystyle{elsarticle-num}
\bibliography{bibliography}

\end{document}